\documentclass[aps,groupedaddress,superscriptaddress,amsmath,amssymb,twocolumn,pra]{revtex4-1}
\usepackage{bm}
\usepackage[dvips]{graphicx}
\usepackage{wrapfig}
\usepackage{subfigure}
\usepackage{stmaryrd}
\usepackage{color}
\usepackage[normalem]{ulem}
\usepackage{enumerate}
\usepackage{epstopdf}
\usepackage{comment}

\def\Pc{P_{\rm cl}}
\def\Pcb{P_{\rm cl}}
\def\Pcc{\bar P}

\def\T{T_{t_F}}
\def\ep{\varepsilon}

\begin{document}
\title{Multiple-period Floquet states and time-translation symmetry breaking\\ in quantum oscillators}

\author{Yaxing Zhang}
\affiliation{Department of Physics, Yale University, New Haven, CT 06511, USA}
\author{J. Gosner}
\affiliation{Institute for Complex Quantum Systems and IQST, Ulm University, 89069 Ulm, Germany}
\author{S. M. Girvin}
\affiliation{Department of Physics, Yale University, New Haven, CT 06511, USA}
\author{J. Ankerhold}
\affiliation{Institute for Complex Quantum Systems and IQST, Ulm University, 89069 Ulm, Germany}
\author{M. Dykman}
\affiliation{Department of Physics and Astronomy, Michigan State University, East Lansing, Michigan 48824, USA}

\begin{abstract}
We study the breaking of the discrete time-translation symmetry in small periodically driven quantum systems. Such systems are intermediate between large closed systems and small dissipative systems, which both display the symmetry breaking, but have qualitatively different dynamics. As a nontrivial example we consider period tripling in a quantum nonlinear oscillator. We show that, for moderately strong driving, the period tripling is robust on an exponentially long time scale, which is further extended by an even weak decoherence. 
\end{abstract}

\date{\today}
\maketitle

The breaking of translation symmetry in time, first proposed by Wilczek \cite{Wilczek2012}, has been attracting much attention recently. Such symmetry breaking can occur only away from thermal equilibrium \cite{Watanabe2015}. It is of particular interest for periodically driven systems, which have a discrete time-translation symmetry imposed by the driving. Here, the time symmetry breaking is manifested in the onset of oscillations with a period that is a multiple of the driving period $t_F$. Oscillations with period $2t_F$ due to simultaneously initialized protected boundary states were studied in photonic quantum walks \cite{Kitagawa2012}; period-two oscillations can also be expected from the coexistence of Floquet Majorana fermions with quasienergies $0$ and $\hbar\pi/t_F$ in a cold-atom system \cite{Jiang2011}. The onset of period-two phases was predicted and analyzed \cite{Khemani2016,Keyserlingk2016,Else2016,Yao2017,Khemani2016a,Bairey2017} in Floquet many-body localized systems, and the first observations of oscillations at multiples of the driving period  in disordered systems were reported \cite{Zhang2016,Choi2016}.  

In systems coupled to a thermal bath, on the other hand,  the effect of period doubling has been well-known. A textbook example is a classical oscillator modulated close to twice its eigenfrequency and displaying vibrations with period $2t_F$ \cite{LL_Mechanics2004}. The oscillator has two states of such vibrations; they have opposite phases, reminiscent of a ferromagnet with two orientations of the magnetization. Several aspects of the dynamics of a parametric oscillator in the quantum regime have been studied theoretically, cf. \cite{Wolinsky1988,Drummond1989,Wielinga1993,Kryuchkyan1996,Marthaler2006,Wustmann2013,Goto2016,Puri2016}, and in experiments, cf. \cite{Nabors1990,Wilson2010,Lin2014}. For a sufficiently strong driving field, a quantum dissipative oscillator, like a classical oscillator, mostly performs vibrations with period $2t_F$. The interplay of quantum fluctuations and dissipation leads to transitions between the period-two vibrational states, but the rate of these transitions is exponentially small \cite{Marthaler2006}.

The goal of this paper is to study time symmetry breaking in isolated or almost isolated driven quantum systems with a few degrees of freedom. They are intermediate between large closed systems and dissipative systems, where the nature of the symmetry breaking is very different. To this end, we analyze a driven nonlinear quantum oscillator. Time symmetry breaking in this system should not be limited to period doubling. As 
an illustration of a behavior qualitatively different from period doubling, we consider period tripling and find the conditions where it occurs. We also address the role of decoherence and the connection between the time symmetry breaking in the coherent and incoherent regimes.

Floquet (quasienergy) states $\psi_\ep(t)$ are eigenstates of the operator $\T$ of time translation by $t_F$, $\T\psi_\ep(t) \equiv \psi_\ep (t+t_F)=\exp(-i\ep t_F/\hbar)\psi_\ep(t)$. For a broken-symmetry state $\psi_{K,\ep_K}$ with $K>1$, time translation by $t_F$ is not described by the factor $\exp(-i\ep t_F/\hbar)$. Instead, $\psi_{K,\ep_K}(t+Kt_F) = \exp(-Ki\ep_K t_F/\hbar)\psi_{K,\ep_K}(t)$. We call $\psi_{K,\ep_K}$ a period-$K$ Floquet state. It is an eigenstate of $T_{Kt_F} = (\T)^K$, but not $T_{t_F}$. 

Multiple-period states naturally occur if the number of states of the system ${\mathbb N}\to \infty$, as in the case of an oscillator. For such systems the quasienergy spectrum is generally dense, cf.~\cite{Hone1997}. Then we can find states $\psi_\ep$ and $\psi_{\ep'}$ with the difference of the quasienergies $|\ep - \ep'|$ infinitesimally close to $\hbar\omega_F/K$ with integer $K{}>1$ (or to $\hbar\omega_F k/K$ with $k<K$); $\omega_F=2\pi/t_F$ is the driving frequency. A linear combination $\alpha \psi_\ep(t) +\alpha'\psi_{\ep'}(t)$ is a period-$K{}$ state. The expectation value of dynamical variables in such a state oscillates with period $Kt_F$. However, the oscillation amplitude will be very small as, generally, the functions $\psi_\ep$ and $\psi_{\ep'}$ will be of a very different form.

\begin{figure}[t]
\includegraphics[scale=0.5]{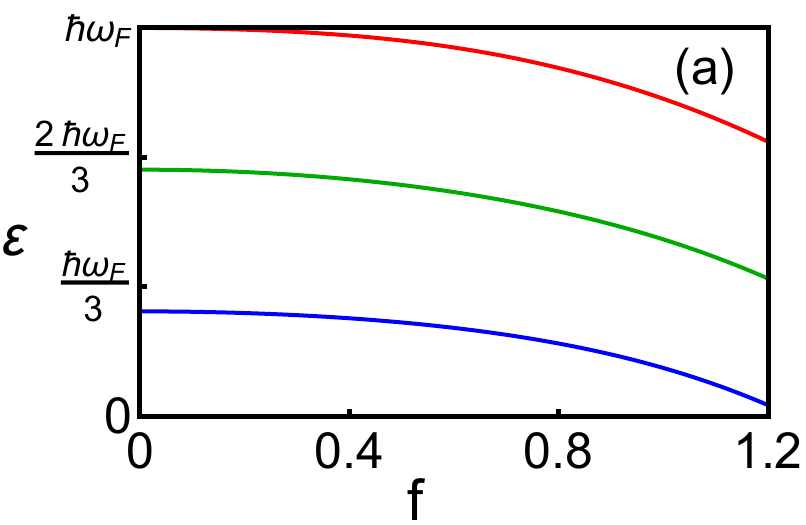}
\hfill
\includegraphics[scale=0.5]{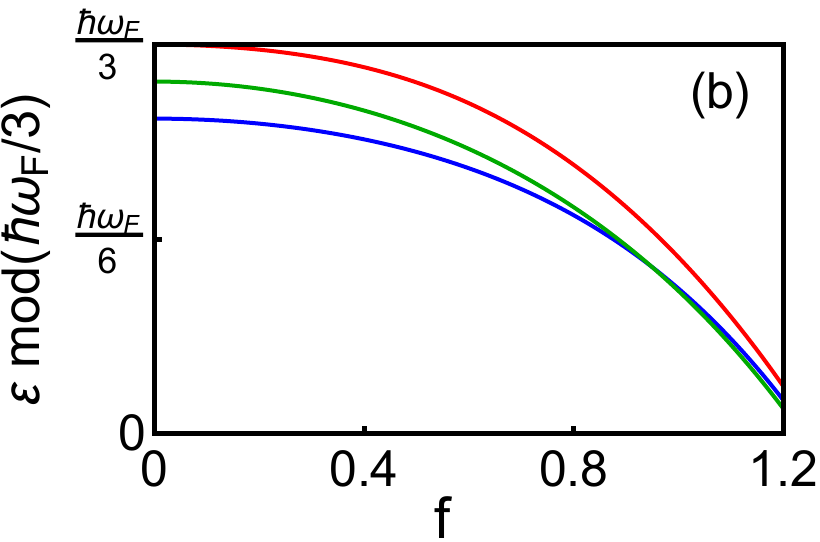}\\
\includegraphics[scale=0.37]{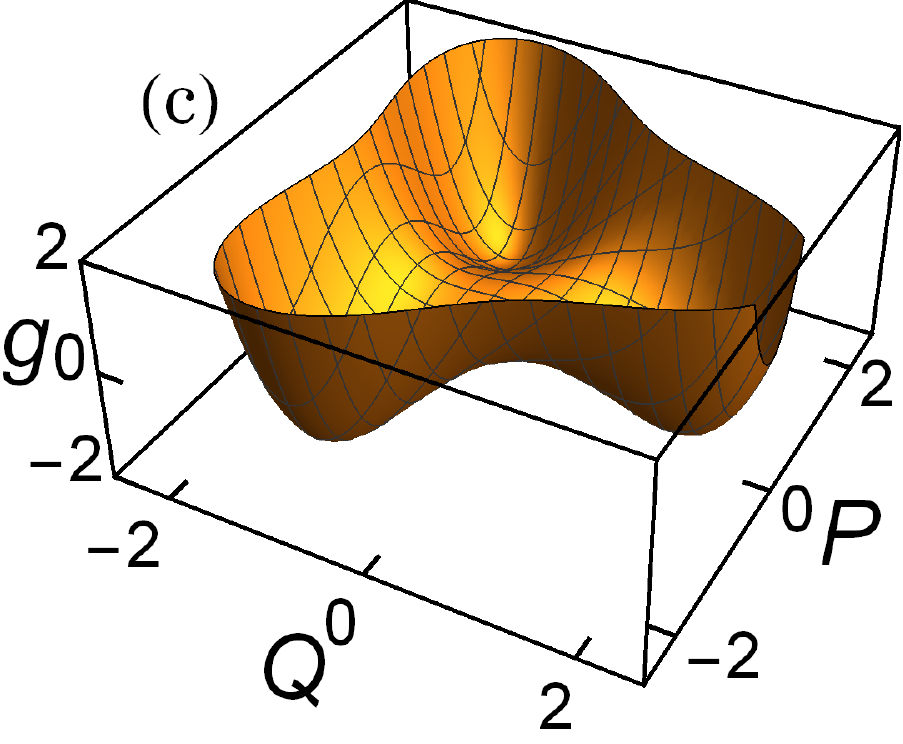}
\hfill
\includegraphics[scale=0.5]{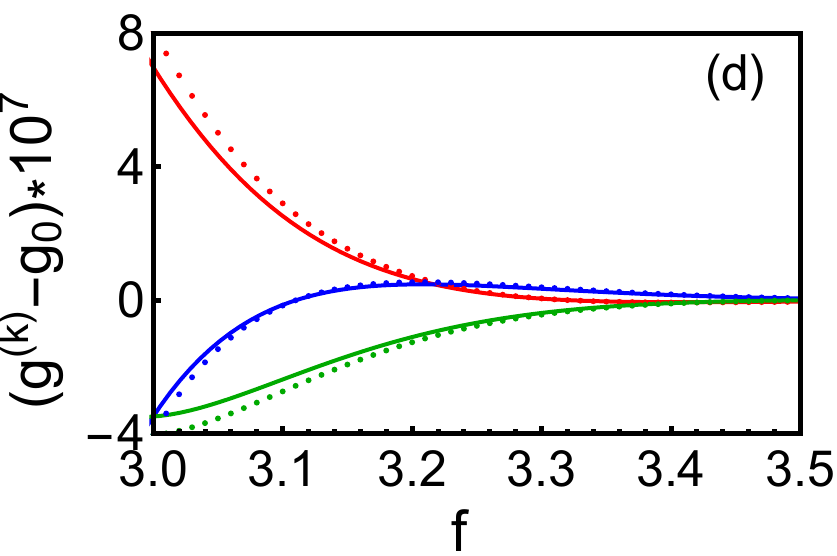}
\caption{(a) Quasienergy levels of strongly overlapping states of a driven oscillator. A period-$K{}$ state occurs when the quasienergy difference equals $\hbar\omega_F/K$. The results are for $K=3$, $f$ is the scaled driving amplitude,  and for $f=0$ the states from top to bottom are the lowest Fock states of the oscillator $|0\rangle, |2\rangle$, and $|1\rangle$. The dimensionless Planck constant for motion in the rotating frame is $\lambda=0.3$ (b) A multiplet formed when the same quasienergies are calculated $\mod(\hbar\omega_F/3)$. (c) The scaled Hamiltonian function $g$ of a nonlinear oscillator driven close to three times the eigenfrequency, Eq.~(\ref{eq:Hamiltonian_RWA}); $Q$ and $P$ are the coordinate and momentum in the rotating frame. (d) Crossing of the quasienergies calculated $\mod(\hbar\omega_F/3)$ for stronger driving; the dotted curves are the analytical results from Eq.~(\ref{eq:splitting_explicit}).}
\label{fig:quasienergy_wrapping}
\end{figure}

The situation is different for an oscillator driven close to an overtone of its eigenfrequency $\omega_0$, i.e., for $\omega_F \approx K\omega_0$. Such an oscillator has several sets of quasienergy states where the quasienergy differences within a set are very close to $\hbar\omega_F/K$ in a broad parameter range, and are exactly equal to $\hbar\omega_F/K$ for some interrelations between the parameters, whereas off-diagonal matrix elements of the dynamical variables are large, see Fig.~\ref{fig:quasienergy_wrapping}. Such states result from tunnel splitting of the  states localized at the minima of the oscillator Hamiltonian in the rotating frame shown in Fig.\ref{fig:quasienergy_wrapping}(c). These localized states correspond to period-$K$ vibrations in the laboratory frame, see below. 

\noindent

In a way, for a parametric oscillator ($K=2$) the occurrence of a period-2 state could be inferred from the results \cite{Marthaler2007a}. However, this state was not identified there and the time symmetry breaking was not addressed. In different terms, sets of states separated by $\approx \hbar\omega_F/K$  were found numerically for $K\gg 1$ for a special model of an oscillator in the interesting paper \cite{Guo2013a}; the considered states did not break time symmetry.  

The period tripling ($K=3$) considered here for a driven oscillator is particularly interesting. It differs from the continuous Landau-type symmetry-breaking transition that occurs for period doubling, cf. \cite{Lin2015}. In the presence of dissipation, the fully-symmetric (zero-amplitude) state does not loose stability. Also, in the quantum regime, there emerges a geometric phase between the broken-symmetry states localized at the minima of the effective Hamiltonian function in phase space, cf. Fig.~\ref{fig:quasienergy_wrapping}(c). Thus, the period-tripling in an oscillator allows one to reveal, using a simple and physically relevant model, the generic conditions for the onset of strongly overlapping multiple-period states and to relate them to the underlying nontrivial symmetry. It also provides a platform for studying quantum tunneling  between localized states in phase space. This problem is considerably different from the classical problem  of tunneling in a symmetric double-well potential \cite{LL_QM81} (see also \cite{Garg2000}). 

We study a most commonly used model of a nonlinear oscillator, the Duffing model, which describes parametric resonance and, as we will see, can describe period tripling; this model refers to a broad range of systems, including trapped relativistic electrons, cold atomic clouds, Josephson junction based systems, and  nanomechanical systems \cite{Tan1991,Dykman2012b}. Its Hamiltonian reads
\begin{align}
\label{eq:hamiltonian}
H=H_0 + H_F, \qquad H_0=\frac{1}{2}p^2 + \frac{1}{2}\omega_0^2 q^2 +\frac{1}{4} \gamma q^4,
\end{align}
where $q$ and $p$ are the oscillator coordinate and momentum. The term $H_F\equiv H_F(t)$ describes the driving. In the analysis of parametric resonance, one chooses $H_F=-\tfrac{1}{2} q^2 F\cos\omega_Ft$ with $\omega_F\approx 2\omega_0$. Here we consider $H_F=-\tfrac{1}{3}q^3F\cos\omega_Ft$ with $\omega_F\approx 3\omega_0$; the results describe also a drive $H_F'=-qF'\cos\omega_Ft$ with $F\to 3\gamma F'/8\omega_0^2$. 

If the driving is not too strong, so that for the states of interest the expectation values of $H_F$ and the nonlinear term $\propto q^4$ are small compared to the harmonic part of $H_0$, the resonant oscillator dynamics can be described in the rotating wave approximation (RWA) \cite{Walls2008}. For an oscillator driven close to the $K{}$th overtone of its eigenfrequency, one makes a canonical transformation $U(t) =\exp(- i a^\dagger a  \omega_F t/K)$, where $a$ and $a^\dagger$ are the ladder operators. 
%
%
The RWA Hamiltonian $H_{\rm RWA}$ is obtained by time-averaging the transformed Hamiltonian $H_K(t) =U^\dagger (t) H(t) U(t) - i\hbar U^\dagger(t)\dot U(t)$, 
\begin{align}
\label{eq:H_RWA_defined}
H_{\rm RWA}=(Kt_F)^{-1}\int_0^{Kt_F}dt H_K(t).
\end{align}
Clearly, $H_{\rm RWA}$ is  independent of time. 

We now establish the relation between the eigenvalues of $H_{\rm RWA}$ and the quasienergies. If $\phi_{}(t)$ is an eigenfunction of $H_{\rm RWA}$, i.e., $H_{\rm RWA}\phi_{} = E\phi_{}$,  then the corresponding wave function in the lab frame is $\psi(t)=U(t)\phi_{}(t)$, and
\begin{align}
\label{eq:define_N_K}
&\T\psi(t) = e^{-iEt_F/\hbar}U(t+t_F)\phi_{}(t)= e^{-i Et_F/\hbar}N_K\psi(t).
\end{align}
We call $E$ the RWA energy. In Eq.~(\ref{eq:define_N_K})
\begin{align}
\label{eq:M_commutator}
N_K = \exp(-2\pi i a^\dagger a/K), \qquad \left[N_K,H_{\rm RWA}\right]=0.
\end{align}
The above commutation relation follows from the relation $H_K(t+t_F) = N_K^\dagger H_K(t)N_K$ and Eq.~(\ref{eq:H_RWA_defined}). Using the explicit form of $H_{\rm RWA}$, the commutation relation (\ref{eq:M_commutator}) was found in Ref.~\onlinecite{Guo2013a} for the same operator as $N_K$.

Operators $N_K^k$ with $k=0,1,...,K-1$ form a cyclic group. Since eigenfunctions of $H_{\rm RWA}$ are also eigenfunctions of $N_K$, one can label them by a superscript $k$, 
\begin{align}
\label{eq:symmetry_property}
N_K\phi^{(k)} = \exp(-2\pi i k_{} /K)\phi^{(k)}, \quad 0\leq k\leq K-1.
\end{align}
Note that $H_{\rm RWA}$ has eigenfunctions with the same $k$, but different $E$. By comparing Eqs.~(\ref{eq:define_N_K}) and (\ref{eq:symmetry_property}) one finds that a wave function $\phi_{}^{(k)}$ with RWA energy $E^{(k)}$ corresponds to a usual Floquet state with quasienergy 
\begin{align}
\label{eq:quasienergy_shift}
\ep^{(k)} = (E^{(k)}+\hbar \omega_F k_{}/K) {\rm mod} (\hbar\omega_F).
\end{align}
As we will see, for sufficiently strong drive the eigenstates of $H_{\rm RWA}$ form multiplets with close eigenvalues $E^{(k)}$ but different $k$. The quasienergies of different states in the multiplets differ by $\approx \hbar\omega_F/K$.

Equation (\ref{eq:symmetry_property}) allows one to write the functions $\phi^{(k)}$ in terms of the Fock states of the oscillator  $|n\rangle$ defined by the condition $a^\dagger a|n\rangle = n|n\rangle$. Only one out of each $K$ Fock states contributes to $\phi^{(k)}$,  $\phi^{(k)} = \sum_nC_n^{(k)}|Kn+k_{}\rangle$. This relation significantly simplifies numerical diagonalization of $H_{\rm RWA}$, as the coefficients $C_n^{(k)}$ with different $k$ are uncoupled. More importantly, it shows that the RWA energy levels of states with different $k$ can {\it cross} when the parameters of the system vary. This crossing is seen in Fig.~\ref{fig:quasienergy_wrapping}. In contrast, the RWA levels of states with the same $k$ {\it avoid crossing}.

The motion in the rotating frame is  conveniently described by the coordinate $Q$  and momentum $P$, which are related to $q$ and $p$ as 
\begin{align}
\label{eq:rotating_frame}
U^\dagger (t)[q+i(K/\omega_F)p]U(t) = C(Q+iP)e^{-i\omega_Ft/K}.
\end{align}
The parameter $C$ is the scaling factor that makes $Q$ and $P$ dimensionless, 
\begin{align}
\label{eq:commutation}
[Q,P]=i\lambda, \qquad \lambda=\hbar K/\omega_FC^2.
\end{align}
The dimensionless Planck constant $\lambda$ and the parameter $C$ in the case of a parametric oscillator, $K=2$, are given in \cite{Marthaler2006}. For the case of period tripling, $C=(8\omega_F\delta\omega/9\gamma)^{1/2}$, where $\delta\omega = \frac{1}{3}\omega_F - \omega_0$ is the frequency detuning from the resonance, $|\delta\omega|\ll \omega_F$. In this case $H_{\rm RWA} = [8\omega_F^2(\delta\omega)^2/27\gamma)]\hat g(Q,-i\lambda \partial_Q)$ with
\begin{align}
\label{eq:Hamiltonian_RWA}
&g(Q,P) = \frac{1}{4}(Q^2+P^2 -1)^2 -\frac{1}{3} f(Q^3 - 3PQP),
\end{align}
where $f=F/(8\omega_F \gamma \delta\omega)^{1/2}$ is the scaled amplitude of the driving. Of interest is the region $\gamma\delta\omega >0$, and we choose $\gamma >0$ and $\delta\omega>0$. 

The function $g(Q,P)$ is the dimensionless Hamiltonian function in the rotating frame. It is plotted in Fig.~\ref{fig:quasienergy_wrapping}. It has a three-fold rotational symmetry in the $(Q,P)$-plane. This symmetry follows from Eqs.~(\ref{eq:M_commutator}) and (\ref{eq:rotating_frame}), since $N_K$ is an operator of rotation by angle $2\pi/K$ in phase plane; the $K$-fold symmetry of $H_{\rm RWA}$ was also seen in \cite{Guo2013a}. 

For moderately strong fields, $g(Q,P)$ has three well-separated minima positioned at the vertices of an equilateral triangle $(Q_m,P_m)$; we count $m=0,1,2$ counterclockwise and
set $m=0$ for the vertex with $P_0=0$.  The eigenstates of the operator $\hat g\equiv g(Q,-i\lambda\partial_Q)$ with the lowest RWA energies are localized near  $(Q_m,P_m)$. In the absence of tunneling, $\hat g$ has three degenerate eigenstates $\Psi_m$. Near their maxima, functions $\Psi_m$ have the form of squeezed ground states of a harmonic oscillator centered at $(Q_m,P_m)$ \footnote{see Supplemental Material for the details of the calculation}. 

The oscillator in a state $\Psi_m$ has a broken time symmetry. The expectation values of dynamical variables oscillate at frequency $\omega_F/3$. Indeed, from Eq.~(\ref{eq:rotating_frame}) time translation by $t_F$ transforms  $\Psi_m\to N_3\Psi_m=\Psi_{m-1}\equiv \Psi_{m+2}$. To come back to state $\Psi_m$, one has to increment time by $3t_F$. The relation $\Psi_{m+1}= N_3^\dagger\Psi_m$ gives the phase shift between functions $\Psi_{m+1}$ and $\Psi_m$. Since $N_3$ is a rotation operator, this phase shift is geometric in nature [34]. 

Tunneling between the minima lifts the degeneracy of the ground state of the operator $\hat g$. In contrast to the problem of tunneling in a symmetric double-well potential \cite{LL_QM81}, $g(Q,P)$ is not even in $Q$, it has three extrema, and two of them lie at nonzero momenta $P$. 

To find the tunnel splitting, we write the wave functions in the coordinate representation, $\Psi_m\equiv \Psi_m(Q)$. The three normalized eigenstates $\phi^{(k)}$ of $\hat g$ with the smallest eigenvalues $g^{(k)}$ ($k=0,1,2$) have the form  
\begin{align}
\label{eq:trial_functions}
&\phi^{(k)}(Q) = \frac{1}{\sqrt{3(1+\delta^{(k)})}}\sum_{m=0,1,2} \Psi_m (Q)e^{-2mk\pi i/3}.
\end{align}
where $\delta^{(k)}= 2{\rm Re}[\langle \Psi_0|\Psi_1\rangle \exp(-2\pi i k/3)]\ll 1$. We choose $\Psi_0(Q)$ to be real and normalized.  Since $\Psi_{m+1} = N_3^\dagger \Psi_m$, we have $\Psi_2(Q)=\Psi_1^*(Q)$. Due to the symmetry,  the functions  $\phi^{(k)}$ can be shown to be orthogonal.

In the spirit of \cite{LL_QM81}, we calculate $g^{(k)}$ using the relation
\begin{align}
\label{eq:splitting_general}
&\int_{\infty}^{Q_*} dQ\Bigl[\phi^{(k)}(Q)(\hat g- g_0)\Psi_0(Q) \Bigr.\nonumber\\
&\Bigl. - \Psi_0(Q)(\hat g -g^{(k)}) \phi^{(k)}(Q)\Bigr]=0
\end{align} 
with $g_0$ being the eigenvalue of $\hat g$ in the state $\Psi_0$, $g_0\approx \min g(Q,P)$ [34].
The difference $g^{(k)}-g_0$ is exponentially small for a small dimensionless Planck constant $\lambda$.

To choose the upper limit $Q_*$ of the integral (\ref{eq:splitting_general}), we note that the functions $\Psi_m(Q)$ fall off exponentially away from the respective $Q_m$, with $\Psi_0$ and $\Psi_{1,2}$ falling off in the opposite directions in the interval $(Q_1,Q_0)$. We choose $Q_*$ within this interval and in such a way that $\Psi_{0,1,2}(Q_*)$ are all of the same order of magnitude and thus can be kept in Eq.~(\ref{eq:trial_functions}) for $\phi^{(k)}(Q)$. The result of integration (\ref{eq:splitting_general}) should be independent of $Q_*$.

The WKB wave functions $\Psi_{0,1}(Q)$ in the classically forbidden region between $Q_1$ and $Q_0$ have the form
\begin{align}
\label{eq:WKB_wave_function}
&\Psi_m(Q)=C_m(i\partial_P g)^{-1/2}e^{iS_m(Q)/\lambda} \quad(m=0,1), \nonumber\\
&\partial_Q S_m = (-1)^m \Pcc(Q), \qquad g(Q,\Pcc) = g_0,
\end{align}
where $S_{0,1}(Q)$ is the  classical action and constants $C_{0,1}$ are found from the matching to the corresponding intrawell wave functions. 

It is critical for understanding the tunneling that, because the effective Hamiltonian function $g(Q,P)$ is quartic in the momentum $P$,  $\Pcc(Q)$ has a branch point $Q_B$ in the interval $(Q_1,Q_0)$. For $Q_1 < Q< Q_B$, $\Pcc(Q)$ has both imaginary and real parts. So does the action $S_m(Q)$. This leads to oscillations of the wave functions in the classically forbidden region. In $S_m(Q)$ one should keep the root with the smallest $|{\rm Im}~\Pcc|$. To describe $\Psi_0$, Eq.~(\ref{eq:WKB_wave_function})  has to be modified  by allowing for a complex conjugate term [34]. 

Calculating the integrals in Eq.~(\ref{eq:splitting_general}) by parts, we find
\begin{align}
\label{eq:splitting_explicit}
&g^{(k)} - g_0 = C_{\rm tun}e^{-S_{\rm tun}/\lambda}\cos (\lambda^{-1}\Phi_{\rm tun}-2\pi k/3),
\end{align}
where $\Phi_{\rm tun} + i S_{\rm tun} = \int_{Q_0}^{Q_1}dQ'\Pc(Q') + P_1Q_1/2 + \lambda G$ with $\Pc$ given by equation $g(Q,\Pc)=\min g(Q,P)$, $G$ being independent of $\lambda$ and having a contribution from the geometric phase, and $C_{\rm tun}\propto \lambda^{1/2}$ [34].

Equation (\ref{eq:splitting_explicit}) shows that the splitting of the eigenvalues of $H_{RWA}$ oscillates as the system parameters vary. Two eigenvalues cross each time $\lambda^{-1}\Phi_{\rm tun} = (n+n'/3)\pi$ with integer $n,n'$. Such crossings are seen in Fig.~\ref{fig:quasienergy_wrapping}. Where the eigenvalues do not cross, they stay exponentially close to each other.

If the oscillator is in a superposition of two states $\phi^{(k)}$ and $\phi^{(k')}$, the expectation values of its variables have period $3t_F$ provided the observation time is smaller than the exponentially long time $|\Omega_{kk'}|^{-1}$, where  the frequency $\Omega_{kk'}=\lambda^{-1}[g^{(k)}-g^{(k')}]\delta\omega$ is determined by the tunnel splitting. The Fourier spectra of the expectation values generally have components at frequency $\omega_F/3 \pm \Omega_{kk'}$; in particular, the coordinate and momentum have just one of these components. This behavior is characteristic also of the oscillator in intrawell states $\Psi_m$, which are superpositions of $\phi^{(1,2,3)}$. The oscillator fluorescence spectrum will display peaks at $\omega_F/3 \pm \Omega_{kk'}$ as well.

It is instructive to compare these results with the period-doubling associated with the topologically protected Floquet boundary states in extended systems \cite{Kitagawa2012,Jiang2011}. To some extent, such states are analogous to the symmetry-protected states $\phi^{(k)}$. If tunneling between the Floquet boundary states can be disregarded, similar to disregarding oscillator tunneling, their combination becomes a  multiple-period state. However, their overlap is exponentially small, in contrast to the functions $\phi^{(k)}$.

The intrawell states $\Psi_m$ are particularly important in the presence of dissipation. Even if the dissipation rate $\Gamma$ is extremely small, but exceeds the exponentially small frequencies  $\Omega_{kk'}$, instead of coherent tunneling between the wells of $g(Q,P)$, the oscillator performs incoherent interwell hopping with typical rate $W<|\Omega_{kk'}|$ [34]. This hopping corresponds to flips of the vibration phase. On times small compared to  $W^{-1}$ the oscillator stays in the multiple-period state inside a well. This is the exact analog of the classical behavior of a dissipative oscillator, including a parametric oscillator, where the multiple-period state is seen on times short compared to the reciprocal rate of interstate switching.

A promising type of oscillator for observing period tripling are modes of microwave cavities coupled to Josephson junctions. Recently there have been studied systems where inelastic Cooper pair tunneling leads to an effective driving of a cavity mode that nonlinearly depends on the mode coordinate and has a tunable frequency $2eV/\hbar$ determined by the voltage $V$ across the Josephson junction \cite{Hofheinz2011,Armour2013,Gramich2013}. There are also other possibilities to resonantly excite multiple-period modes in microwave cavities \footnote{P. Delsing, D. Esteve, and  F. Portier, private communications}.

In conclusion, we studied a quantum oscillator driven close to an overtone of its eigenfrequency and showed that a small quantum system can display coherent multiple-period dynamics. We explicitly described this dynamics for the previously unexplored nontrivial case of period tripling and established the relation to protected boundary Floquet states in extended systems and to multiple-period states in dissipative systems. 

We are grateful to G. Refael, M. Rudner, and S. Sondhi for the discussions and correspondence.  YZ and SMG were supported by the U.S. Army Research Office (W911NF1410011) and by the National Science Foundation (DMR-1609326).;  JG and JA were supported in part by the German Science Foundation through SFB/TRR 21 and the Center for Integrated Quantum Science and Technology (IQST);  MID was supported in part by the National Science Foundation (Grant No. DMR-1514591).

\bibliographystyle{apsrev4-1}
%
\newpage

\newpage

\def\Pc{\bar P}

\begin{center}
{\Large\bf  Supplemental Material}
\end{center}
\section{The Intrawell Wave Functions of the RWA Hamiltonian}

We consider the dynamics of the oscillator driven close to three times its eigenfrequency in the rotating wave approximation (RWA). The scaled RWA Hamiltonian function $g(Q,P)$, which is given by Eq.~(10) of the main text and is plotted there in Fig. 1, has three symmetrically located minima at  points $(Q_m,P_m)$ with $m=0,1,2$,
\begin{align}
\label{eq:equilibrium_positions}
&Q_0=\frac{1}{2}\left[f+(f^2+4)^{1/2}\right], \qquad Q_1 = Q_2 = -Q_0/2,\nonumber\\
&P_0=0,\qquad P_1 = -P_2 = \sqrt{3} Q_0/2, 
\end{align}
The minimal value $g_{\min}$ of $g(Q,P)$ and the dimensionless frequency of classical vibrations about a minimum  $\omega_{\min} = \left(\det[\partial^2_{x_i x_j} g(x_1,x_2)]\right)^{1/2}$ (the derivatives are calculated at a minimum of $g$) are
\begin{align}
\label{eq:gmin}
g_{\min}= -\frac{1}{12}fQ_0(Q_0^2+3),\quad   \omega_{\min}= [3fQ_0(Q_0^2 +1)]^{1/2}\, .
\end{align}
The frequency $\omega_{\min}$ is the same for all minima. So is also the lowest eigenvalue $g_0$ of the Hamiltonian $\hat g(Q,-i\lambda\partial_Q)$ in the neglect of tunneling. To the lowest order in the dimensionless Planck constant $\lambda$ it corresponds to the lowest eigenvalue of a harmonic oscillator, \[g_0=g_{\min}+ \tfrac{1}{2}\lambda \omega_{\min}.\]

The calculation of the tunnel splitting is done below by first finding the intrawell wave functions $\Psi_m(Q)$ near their maxima inside the well, then finding the geometric phase shift between different $\Psi_m$, and then explicitly writing down the WKB tails of functions $\Psi_m$ in the classically forbidden regions, which are given by Eq.~(13) of the main text. Since $\Psi_2(Q) = \Psi_1^*(Q)$, we only need to find $\Psi_0(Q)$ and $\Psi_1(Q)$.

\subsection{The wave function $\Psi_0(Q)$}
Near the minimum $(Q_0, P_0)$ we have $g(Q,P)\approx g_{\min}+\frac{1}{2}(Q_0^2+1)(Q-Q_0)^2 + \frac{3}{2}fQ_0P^2$.
The wave function $\Psi_0(Q)$ is Gaussian for $|Q-Q_0|\ll |Q_1-Q_0|$ and can be chosen to be real,
\begin{equation}
\label{eq:psi_0_intrawell}
\Psi_0(Q) = (\sqrt{\pi} l_q)^{-1/2}\exp[-(Q-Q_0)^2/2l_q^2],
\end{equation}
with $l_q=[\lambda\omega_{\min}/(Q_0^2+1)]^{1/2}$ being the localization length.

We are interested in the tail of $\Psi_0$ for $Q$ between the minima of $g(Q,P)$, i.e., for $Q_1 < Q < Q_0-l_q$. The WKB form of $\Psi_0(Q)$ is given by Eq.~(13) of the main text, which we here write explicitly,
\begin{align}
\label{eq:WKB_wave_function_1}
&\Psi_0(Q)=C_0(i\partial_P g)^{-1/2}\exp[iS_0(Q)/\lambda], \nonumber\\
 &S_0(Q)=\int_{Q_0-l_q}^Q dQ'\Pc(Q'), 
\end{align}
where $\Pc(Q)$ is given by equation $g(Q,\Pc) = g_0$ and $\partial_Pg$ is calculated for $P=\Pc(Q)$. For the branch of $\Pc$ that we are interested in
\begin{align}
\label{eq:momentum_general}
&\Pc(Q)^2= A(Q) + B^{1/2}(Q), \qquad A(Q)= 1- Q^2 -2fQ, \nonumber\\
&B(Q) = A^2(Q) -4 [g(Q,0)-g_0],
\end{align}
with Im~$\Pc <0$ for $Q<Q_0$; we keep the correction $\propto \lambda$ to secure matching to Eq.~(\ref{eq:psi_0_intrawell}).

For $Q$ close to $Q_0$ and $Q<Q_0-l_q$, we have $A(Q)<0, B(Q)>0$, and $A(Q)+ B^{1/2}(Q) <0$. Therefore $\Pc(Q)$ is purely imaginary and the same is true for the function
\begin{align}
\label{eq:dgdp}
\partial_Pg = \Pc(Q)B^{1/2}(Q)
\end{align}
 with $i\partial_Pg >0$. Accordingly, $\Psi_0(Q)$ exponentially decays with increasing $Q_0-Q$. The prefactor $C_0$ is determined by matching Eqs.~(\ref{eq:psi_0_intrawell}) and (\ref{eq:WKB_wave_function_1}) for $Q$ close to $Q_0$ but $Q_0-Q\gg l_q$,
\[C_0=(\omega_{\min}/2\sqrt{\pi e})^{1/2}.\]

As $Q$ decreases, first $B(Q)$ becomes equal to zero at point $Q_B$. To the leading order in $\lambda \ll 1$
\begin{align}
\label{eq:QB}
Q_B \approx Q_0-\frac{3}{4}f.
\end{align}
For still smaller $Q$,  $A(Q)$ changes sign to positive. This happens for  $Q_B> Q>Q_1\equiv -Q_0/2$. Importantly,
\begin{align}
\label{eq:-Q0over2}
A(Q_1)= P_1^2 >0, \quad  B(Q_1) =2\lambda\omega_{\min}\, .
\end{align}

In the explicit form, the imaginary part of the momentum in the classically forbidden region is
\begin{align}
\label{eq:imaginary_P}
&{\rm Im}\Pc(Q)=-\left[-A(Q)-B^{1/2}(Q)\right]^{1/2} \quad(Q_B<Q<Q_0)\nonumber\\
&{\rm Im}\Pc(Q)= -\left[(A^2+|B|)^{1/2}-A\right]^{1/2}/\sqrt{2}  \quad(Q<Q_B).
\end{align}

As discussed in the main text, the level splitting crucially depends on the oscillations of the wave function under the barrier. These oscillations start with the decreasing $Q$ at $Q=Q_B$. Near $Q_B$ we have $B(Q)\approx\partial_Q B(Q_B)(Q-Q_B)$, whereas $A(Q_B)<0$. Therefore  $\Pc \approx -i|A(Q_B)|^{1/2} + (i/2)|\partial_QB(Q_B)/A(Q_B)|^{1/2}(Q-Q_B)^{1/2}$ for small $Q-Q_B>0$, i.e., $Q_B$ is a branching point of $\Pc(Q)$. We have to go around above and below this point in the complex plane to obtain the wave function for $Q<Q_B$, following the standard procedure \cite{LL_QM81_1}. As a result, we find for $Q<Q_B$
\begin{align}
\label{eq:transmitted_psi}
&\Psi_0(Q)\approx 2C_0|\partial_Pg|^{-1/2}\exp[-{\rm Im}~S_0(Q)/\lambda]\cos \Phi_0(Q),\nonumber\\
&\Phi_0(Q) = \Phi_0'(Q)+\Phi''_0(Q).
\end{align}
Here, the phase $\Phi'_0(Q)$ comes from the real part of the action,
\begin{align}
\label{eq:Phi0prime}
&\Phi_0'(Q) = \lambda^{-1}\int_{Q_B}^Q dQ'\, {\rm Re}\Pc(Q'),\nonumber\\
&{\rm Re} \Pc(Q) = -\left[(A^2+|B|)^{1/2}+A\right]^{1/2}/\sqrt{2},
\end{align}
whereas $\Phi_0''(Q)$ comes from the prefactor, with account taken of going around $Q_B$ in the complex plane,
\begin{align}
\label{eq:Phi0_dblprime}
\Phi_0''(Q) = -\frac{1}{2}\arcsin\left[{\rm Re}\Pc(Q)/|\Pc(Q)|\right] - \frac{\pi}{4}.
\end{align}
The choice of ${\rm Re}\Pc$ and ${\rm Im}\Pc$ in Eqs.~(\ref{eq:imaginary_P}) and (\ref{eq:Phi0prime}) corresponds to writing $B^{1/2} = i|B|^{1/2}$ in Eq.~(\ref{eq:momentum_general}) for $\Pc^2$ in the region where $B(Q)<0$.

The WKB approximation (\ref{eq:WKB_wave_function_1}) breaks down near $Q_1$, as $B(Q)$ becomes $\sim \lambda$ and $|\partial_Pg|$ becomes small. However, we do not need to calculate the wave function $\Psi_0(Q)$ in this region, as seen from Eq.~(12) of the main text.

\subsection{The wave function $\Psi_1(Q)$}

The minimum of $g(Q,P)$ at $(Q_1,P_1)$ corresponds to a nonzero momentum $P_1>0$. Therefore the wave function $\Psi_1$ centered at $Q_1$ is complex valued even near its maximum. Calculating $\Psi_1$ involves three steps: finding it inside the well of $g(Q,P)$ near $Q_1,P_1$; finding the geometric phase, that relates $\Psi_1$ and $\Psi_0$ given that $\Psi_0$ is chosen in the form (\ref{eq:psi_0_intrawell}), and then finding the tail of $\Psi_1$ in the classically forbidden range.

\subsubsection{The intra-well wave function and the geometric phase}

Using the explicit form (\ref{eq:equilibrium_positions}) of $Q_1,P_1$, to the second order in $\delta Q = Q-Q_1, \delta P=P-P_1$ we write the Hamiltonian near $(Q_1,P_1)$ as
\begin{align}
\label{eq:g_near_upper_P}
g(Q,P)\approx g_{\min} &+ \frac{3}{4}(1+fQ_0)\delta P^2 +\frac{1}{4}(1+5fQ_0)\delta Q^2\nonumber\\
& +(\sqrt{3}/4)(fQ_0-1)[\delta Q \delta P +{\rm h.c.}].
\end{align}
The expression for $\Psi_1$ for $|\delta Q|\ll Q_0-Q_1$ then reads
\begin{align}
\label{eq:psi1_intrawell}
&\Psi_1(Q)=C_{1,{\rm intra}}\exp[(iP_1\delta Q- \frac{1}{2}\beta \delta Q^2)/\lambda],\nonumber\\
&\beta =[2\omega_{\min}  +i \sqrt{3}(fQ_0-1)]/3Q_0^2.
\end{align}
The Gaussian-width parameter $\beta$  is now complex-valued. So is also the prefactor $C_{1,{\rm intra}}$, which has a phase factor $\exp(i\theta_1)$.

The phase $\theta_1$ has a geometric nature. It is determined by the fact that, as indicated in the main text, $\Psi_1$ and $\Psi_0$ are related by the transformation of rotation in phase plane, $\Psi_1=N_3^\dagger\, \Psi_0$. Here, $N_3 = \exp(-2\pi ia^\dagger a/3)$ with $a = (2\lambda)^{-1/2}(Q+ iP)\equiv (2\lambda)^{-1/2}(Q+ \lambda\partial_Q)$. To calculate $\theta_1$, we consider a coherent state in the coordinate representation
\[|\alpha\rangle = \frac{1}{(\pi\lambda)^{1/4}}\exp\left\{-\frac{1}{2}(|\alpha|^2-\alpha^2) -
\frac{[Q-(2\lambda)^{1/2}\alpha]^2}{2\lambda}\right\}\]
and set $\alpha =Q_0/\sqrt{2\lambda}$, so that the wave function $|\alpha\rangle$ is centered at $Q_0$ and thus  strongly overlaps with $\Psi_0$. Since the function $\Psi_1$ is obtained from $\Psi_0$ by applying to $\Psi_0$ the operator $N_3^\dagger$,  we can write the overlap integral as $\langle \alpha|\Psi_0\rangle =\langle \alpha|N_3\Psi_1\rangle= \langle \alpha \exp(2\pi i/3)|\Psi_1\rangle$. The ``rotated" state $|\alpha \exp(2\pi i/3)\rangle$ strongly overlaps with $\Psi_1$. Therefore the above overlap integrals can be calculated using the explicit Gaussian form of $\Psi_0$ and  $\Psi_1$ near their maxima. With account taken of the normalization of $\Psi_1$, this gives
\begin{align}
\label{eq:phase_psi1}
&C_{1,{\rm intra}}=[{\rm Re} \beta/\pi\lambda]^{1/4}\exp(i\theta_1),\nonumber\\
&\theta_1 = \frac{1}{2}\arg (\beta +1)  +P_1Q_1/2\lambda.
\end{align}

\subsubsection{The wave function $\Psi_1$ in the classically forbidden region}

It is clear from Eq.~(12) of the main text that we need to find the tail  of the wave function $\Psi_1$ in the classically forbidden region only for $ Q > Q_1$. It is given by Eq.~(13) of the main text. In a more explicit form
\begin{align}
\label{eq:WKB_psi1}
&\Psi_1(Q)=C_1(i\partial_P g)^{-1/2}\exp[iS_1(Q)/\lambda], \\
 &S_1(Q)=-\int_{Q_1 + l_q'}^Q dQ'\,\Pc(Q'),\nonumber
\end{align}
where $\Pc(Q)$ is given by Eqs.~(\ref{eq:imaginary_P}) and (\ref{eq:Phi0prime}), $l_q' =[\lambda/{\rm Re}\, \beta]^{1/2}$ . Equation (\ref{eq:WKB_psi1}) corresponds to choosing $B^{1/2}(Q) = i|B(Q)|^{1/2}$ for $B(Q) < 0$ and to $\partial_Pg$ calculated for $P(Q)=\Pc(Q)$, i.e., $\partial_Pg=\Pc B^{1/2}$.  For $Q_B-Q\gg Q-Q_1\gg l'_q$ we have $-\Pc(Q) \approx P_1+ i\beta(Q-Q_1)$, as expected from Eq.~(\ref{eq:psi1_intrawell}). By matching Eqs.~(\ref{eq:psi1_intrawell}) and (\ref{eq:WKB_psi1}),  we find
\begin{align}
\label{eq:C_1}
&C_1=(\omega_{\min}/2\sqrt{\pi e})^{1/2} \exp(i\theta_1'),\nonumber\\
&\theta_1' = \theta_1 -\lambda^{-1}\left[ (l_q^{\prime\,2}/2){\rm Im}\beta-P_1l_q' \right].
\end{align}
Because we count the action $S_1$ off from $Q_1 + l_q'$, there emerges an extra phase factor in $C_1$ due to the oscillations of the wave function inside the ``potential well" centered at $(Q_1,P_1)$.

\section{Tunnel splitting of the scaled RWA energy levels}

The scaled RWA energies $g^{(k)}$ give the values of the quasienergies $\ep^{(k)}$ of the driven oscillator, $\ep^{(k)} = (\Xi g^{(k)} + \hbar\omega_F k/3)\!\!\!\mod(\hbar\omega_F)$, where $\Xi = |8\omega_F^2(\delta\omega)^2/27\gamma|$, see Eqs.~(6) and the text above Eq. (9) of the main text. 
The explicit expressions for the wave functions (\ref{eq:transmitted_psi}) and  (\ref{eq:WKB_psi1}) allow us to calculate the scaled energies $g^{(k)}$ using  Eq.~(12) of the main text, which we reproduce here for convenience,
\begin{align}
\label{eq:splitting_general_1}
&\int_{\infty}^{Q_*} dQ\Bigl[\phi^{(k)}(Q)(\hat g- g_0)\Psi_0(Q) \Bigr.\nonumber\\
&\Bigl. - \Psi_0(Q)(\hat g -g^{(k)}) \phi^{(k)}(Q)\Bigr]=0,
\end{align}
Functions $\phi^{(k)}$ are sums of functions $\Psi_m(Q)$ weighted with factors $\exp(-2\pi i mk/3)/\sqrt{3}$. For $Q_*$ well inside the interval $(Q_1,Q_0)$ , we have $ \int_\infty^{Q_*}\Psi_0^2(Q)dQ =-1$. Taking into account that overlapping of the functions $\Psi_{1,2}(Q)$ with $\Psi_0(Q)$ is exponentially small, we rewrite Eq.~(\ref{eq:splitting_general_1}) as
\begin{align}
\label{eq:splitting_implicit}
 g^{(k)}-g_0 \approx &\left[ \int_\infty ^{Q_*}dQ \,\Psi_1(Q)\hat g \Psi_0 - \int_\infty ^{Q_*}dQ \,\Psi_0\hat g \Psi_1(Q)\right] \nonumber\\
&\times \exp(-2k\pi i/3) + {\rm c.c.}
\end{align}
 It is important that the product $\Psi_0(Q)\Psi_1(Q)$ has two terms. One of them is $\propto \exp\{i[S_0(Q) + S_1(Q)]\}$. It smoothly depends on $Q$, because $S_0(Q) + S_1(Q) ={\rm const}$ for $Q_1 < Q < Q_0$. The other term is  $\propto \exp\{-i[S_0^*(Q) - S_1(Q)]\}$, it is a fast oscillating function of $Q$. The contribution of this term to the integrals (\ref{eq:splitting_implicit})  is exponentially small and exponentially sensitive to the change of $Q_*$ on the scale $\propto \lambda$. Therefore this term  should be disregarded.

Using the explicit form of the operator $g(Q, -i\lambda\partial_Q)$ and integrating by parts, from Eq.~(\ref{eq:splitting_implicit}) we obtain
\begin{align}
\label{eq:splitting_more_halfbaked}
&g^{(k)} - g_0 =- 2\lambda C_0|C_1|\exp(-S_\lambda/\lambda)\cos\left(\frac{\Phi_\lambda}{\lambda}-\frac{2k\pi}{3}\right),\nonumber\\
&S_\lambda =  -\int_{Q_1+l_q'}^{Q_0-l_q}dQ \,{\rm Im}\Pc(Q),\nonumber\\
&\Phi^{(k)}_\lambda = -\int_{Q_1 + l_q'}^{Q_B}dQ \,{\rm Re}\Pc(Q)  +\lambda\theta_1'.
\end{align}

This expression is somewhat inconvenient, as $\Pc$ is calculated with account taken of the term $\propto \lambda$. It is easy to see that $\Pc(Q) \approx {}\Pcb(Q) +\frac{1}{2}\lambda\omega_{\min}/\partial_Pg$, where $\Pcb$ is given by the value of $\Pc$ calculated for $\lambda=0$. This approximation breaks down near $Q_0, Q_B$ and $Q_1$ where $\partial_Pg$ goes to zero. Similar to Ref.~\onlinecite{Garg2000_1},  for $Q_0>Q>Q_B$ one can write
\begin{align}
\label{eq:intermediate1}
&\int_{Q_0-l_q}^QdQ' \,\Pc(Q')\approx \int_{Q_0}^Q dQ'\left[\Pcb(Q') + \lambda Y(Q',Q_0) \right]\nonumber\\
& -\frac{i\lambda}{2}\log\frac{|Q-Q_0|}{l_q} -\frac{i\lambda}{4}-\frac{i\lambda}{2}\log 2,\nonumber\\
& Y(Q,Q_m)= \frac{\omega_{\min}}{2{}\Pcb(Q){}B_{\rm cl}^{1/2}(Q)} -\frac{i}{2|Q-Q_m|}\, .
\end{align}
Here, ${}B_{\rm cl}(Q)= (16f/3)(Q-Q_1)^2(Q- Q_B)$ is the value of $B(Q)$ calculated for $\lambda=0$. A similar transformation can be made for $\int_{Q_1+ l_q'}^Q dQ'\Pc(Q')$ in the region $Q_1<Q<Q_B$.

We now have to consider the vicinity of $Q_B$. Formally, the quantum correction to $\Pc$ diverges at $Q_B$. However, the divergence is integrable. Therefore Eq.~(\ref{eq:intermediate1}) applies all the way till $Q=Q_B$, and one can use the value of $Q_B$ given by Eq.~(\ref{eq:QB}).

The final result for the difference of the scaled RWA energies is Eq.~(14) of the main text,
\begin{align}
\label{eq:splitting_explicit_1}
g^{(k)} - g_0 = C_{\rm tun}\, e^{-S_{\rm tun}/\lambda}\cos\left(\lambda^{-1}\Phi_{\rm tun} -2\pi k/3\right)\, ,
\end{align}
with real $S_{\rm tun}$ and $\Phi_{\rm tun}$,
\begin{align}
\label{eq:S_tunnel_explicit}
&S_{\rm tun} = \int_{Q_0}^{Q_1}dQ\,  {\rm Im} {}\Pcb(Q) + \lambda \,{\rm Im}\, K_{\rm tun} \nonumber\\
&\Phi_{\rm tun}= \int_{{} Q_B}^{Q_1}dQ \,{\rm Re}{}\Pcb(Q)  + \lambda{\rm Re}\, K_{\rm tun} +\lambda\theta_1,\nonumber\\
&C_{\rm tun}=-\frac{3}{2}\sqrt{\lambda}\,\omega_{\min}\left[\frac{2(Q_0^2 +1)}{3\pi^2 Q_0^2}\right]^{1/4}
\left[f(2Q_0-f)\right]^{1/2}.
 \end{align}
Here,
\begin{align}
\label{eq:K_tun}
 &K_{\rm tun} = \int_{Q_0}^{{} Q_B}dQ \,Y(Q,Q_0) + \int_{{} Q_B}^{Q_1}dQ\, Y(Q,Q_1)
\end{align}
and $\theta_1$ is given in (\ref{eq:phase_psi1}).

\begin{figure}[h]
\begin{center}
\includegraphics[scale=0.81]{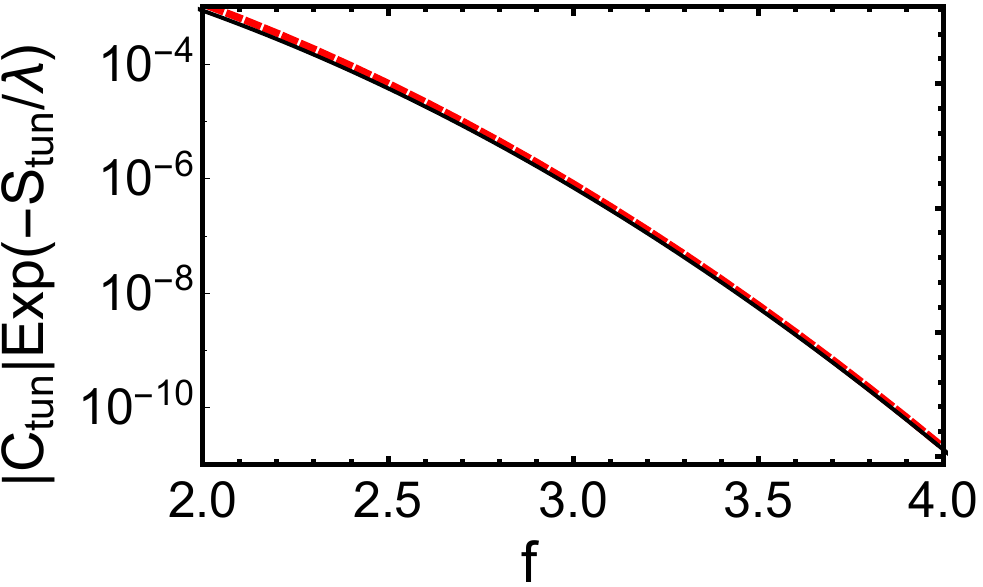} \\
\includegraphics[scale=0.74]{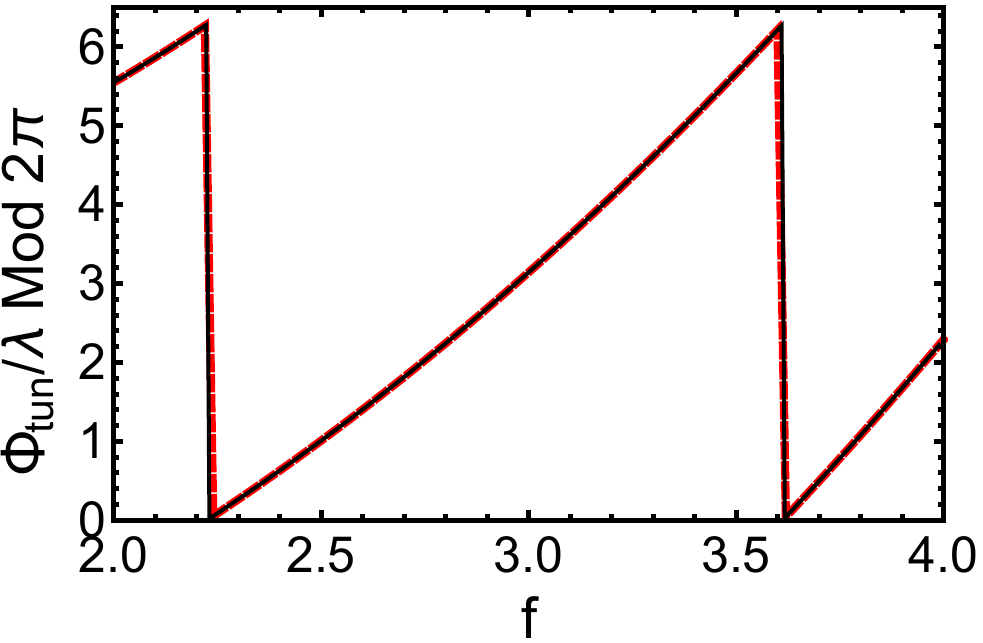}
\end{center}
\caption{Upper pane: the amplitude of the tunnel splitting of the scaled RWA energy levels, which is given by Eq.~(\ref{eq:splitting_explicit_1}) without the oscillating factor. Lower panel: the phase of the tunnel splitting. Black solid lines:  the results of the numerical solution of the eigenvalue problem for the Hamiltonian $g(Q,-i\lambda\partial_Q)$; red dashed lines: Eq.~(\ref{eq:splitting_explicit_1}) The results refer to $\lambda = 0.3$.}
\label{fig:detailed_comparison}
\end{figure}

The explicit expression (\ref{eq:splitting_explicit_1}) is in an extremely good agreement with the numerical calculations. This can  be seen from Fig.~1 in the main text. A more detailed comparison is shown in Fig.~\ref{fig:detailed_comparison}.  Equation (\ref{eq:splitting_explicit_1}) simplifies in the limit of comparatively strong drive, $f\gg 1$. The leading order terms in $S_{\rm tun}$ and in $\Phi_{\rm tun}$ are quadratic in $f$. Numerically, the asymptotic regime is reached for comparatively large $f$, where the tunneling amplitude becomes very small.

\section{Quantum Diffusion over the broken-symmetry states}

The dynamics of the driven oscillator system can be strongly changed by an already very weak dissipation. Two types of dissipative processes can be conditionally separated. One of them causes transitions between the states that belong to the same multiplet  formed by the tunnel splitting of a quantized state of motion inside  a well of $g(Q,P)$. In particular, in this paper we considered such multiplet $\phi^{(k)}$ formed by the tunnel splitting of the lowest quantized intrawell state. The other dissipative process leads to transitions between the intrawell states. 

In terms of the dissipation mechanisms, an important type of physical  dissipative processes are transitions between the Fock states of the oscillator with emission or absorption of excitations of the thermal reservoir to which the oscillator is coupled. Another mechanism is fluctuations of the oscillator eigenfrequency due to the coupling to a reservoir or due to an external noise. It leads to dephasing of the vibrations, but not to an appreciable energy exchange with the reservoir. There may be also dissipation channels that are induced by the driving field; however, for the considered comparatively weak resonant field they are not important. 

We note first that the dephasing does not mix the states within the tunnel-split multiplets. Indeed, as indicated in the main text, the wave functions $\phi^{(k)}$ can be written in terms of the Fock states of the oscillator $|n\rangle$ as $\phi^{(k)} = \sum_nC_n^{(k)}|3n+k\rangle$. The coupling to a thermal bath, which leads to dephasing, has the form $H^{\rm (ph)} = a^\dagger a H_b^{\rm (ph)}$, where $a,a^\dagger$ are the oscillator ladder operators and $ H_b^{\rm (ph)}$ is an operator that depends on the dynamical variables of the bath only. Clearly, such coupling is diagonal in the $\phi^{(k)}$ basis.

The simplest coupling that leads to the oscillator energy relaxation is linear in $a,a^\dagger$. To the lowest order of the perturbation theory, for the well-understood conditions, it is described by the term $\dot \rho_d$ in the equation for the oscillator density matrix $\rho$ in slow time compared to $\omega_F^{-1}$, 
\begin{align}
\label{eq:dissipative}
&\dot\rho_d = -\Gamma\left[(\bar n +1) (a^\dagger a\rho -2 a \rho a^\dagger + \rho a^\dagger a) \right.\nonumber\\
&\left. +\bar n (a a^\dagger \rho -2 a^\dagger \rho a + \rho aa^\dagger) \right],
\end{align}
where $2\Gamma$ is the energy decay rate of the oscillator and $\bar n=[\exp(\hbar\omega_0/k_BT) -1]^{-1}$ is the oscillator Planck number. In what follows we assume that $\bar n=0$; an extension to a nonzero Planck number is straightforward and does not affect the result.

The goal of this section is to show that, even where $\Gamma$ is extremely small, but exceeds the tunnelling frequency $\Omega_{kk'} =\lambda^{-1}\delta\omega (g^{(k)} - g^{(k')})$, the oscillator dynamics changes qualitatively compared to the coherent dynamics. Instead of coherent tunneling between the intrawell states $\Psi_m$, which have broken time translation symmetry, the oscillator performs random hopping between the wells.

We first discuss the effect of the dissipation (\ref{eq:dissipative}) by disregarding the dissipation-induced transitions between the intrawell states. In this approximation, one can describe the evolution of the oscillator in terms of the kinetic equation for the matrix elements $\rho_{mm'} \equiv \langle \Psi_m|\rho|\Psi_{m'}\rangle$. The interwell tunneling can be mapped onto the tight-binding model with Hamiltonian 
\begin{align}
\label{eq:tunnel_site}
H_{\rm tun} =t_{\rm tun}\sum_{m=0,1,2} |\Psi_m\rangle\langle \Psi_{m+1}| + {\rm H.c.},
\end{align}
where we use the convention $|\Psi_3\rangle \equiv |\Psi_0\rangle$. The hopping integral is $t_{\rm tun}= (\hbar\delta\omega/2\lambda)C_{\rm tun}\exp[(-S_{\rm tun}+i\Phi_{\rm tun})/\lambda]$ with $C_{\rm tun}, S_{\rm tun}$, and $\Phi_{\rm tun}$ given by Eq.~(\ref{eq:S_tunnel_explicit}).

To the leading order in $\lambda$, we have $\langle\Psi_m|a|\Psi_{m'}\rangle = (2\lambda)^{-1/2}(Q_m+iP_m)\delta_{mm'}$. Therefore, from Eq.~(\ref{eq:dissipative}), off-diagonal matrix elements $\rho_{mm'}$ decay with rate $\propto \Gamma/\lambda$. If this rate exceeds $|\Omega_{kk'}| \sim |t_{\rm tun}|/\hbar$, then over time $\sim \lambda/\Gamma$ the off-diagonal matrix elements decay to their quasi-stationary values, which are determined by the diagonal matrix elements $\rho_{mm}$. The latter vary much slower, 
\begin{align}
\label{eq:q_diffusion}
&\dot \rho_{mm} = W\sum_{m'\neq m} \rho_{m'm'}- 2W\rho_{mm}, \nonumber\\
& W=\lambda |t_{\rm tun}|^2/\hbar^2 \Gamma Q_0^2.
\end{align}
Parameter $W$ is the rate of hopping between the wells of $g(Q,P)$, it is much smaller than the tunneling frequency $|\Omega_{kk'}|$. The hopping is a Poisson process in the slow time, it is incoherent and is a discrete analog of diffusion. The above analysis is in the spirit of the theory of quantum diffusion in solids \cite{Kagan1992} and its analog in systems with a small number of potential wells \cite{Dykman1978b}.

The role of the dissipation-induced intrawell transitions is more subtle. Even for $T=0$, these transitions lead to an occupation of excited intrawell states, cf.~\cite{Marthaler2006_1}. On the time scale determined by $1/\Gamma$, near the minimum of a well there is progressively formed a Boltzmann-type distribution over the states. The stationary ratio of the populations of the neighboring states can be shown to be $(1+2fQ_0-\omega_{\min})/(1+2fQ_0+\omega_{\min})$. 

The  tunnel splitting increases for higher-lying intrawell states. However, near the minimum of $g(Q,P)$ this increase is slow. The tunneling action $S_{\rm tun}(n)$ varies with the intrawell level number $n$ as $|\partial S_{\rm tun}(n)/\partial n| =\lambda\omega_{\min}\tau_n$. Here, $\tau_n$ is the dimensionless imaginary time of interwell tunneling given by Im~$\int dQ/\partial_Pg$, where the classical momentum is calculated  for $g(Q,P) = g_{\min} +\lambda\omega_{\min}(n+1/2)$. This time is logarithmically large for small $n$. Therefore, for small $\Gamma$ but still $\Gamma \gg |t_{\rm tun}|/\hbar$, tunneling via excited intrawell states weakly renormalizes the rate $W$ in Eq.~(\ref{eq:q_diffusion}). 

Even if for highly excited intrawell states, with $n$ exceeding some critical $n_{\rm cr}\propto 1/\lambda$, the hopping integral exceeds $\Gamma/\hbar$, interwell switching via these states will be very slow, as the occupation of these states will be small. We note that, if $\hbar \Gamma$ exceeds the hopping integral for almost all intrawell states, interwell switching may occur via dissipation-induced transitions over the interwell barrier of $g(Q,P)$, i.e., over the saddle point of $g(Q,P)$ seen in Fig.~1 of the main text. This is the dominating switching mechanism for a parametric oscillator \cite{Marthaler2006_1}.

\bibliographystyle{apsrev4-1}


%

\end{document}